\documentclass[twocolumn,showpacs,preprintnumbers,amsmath,amssymb,superscriptaddress,UTF8]{revtex4}
\usepackage{dsfont}
\usepackage{cases}
\usepackage{graphicx}% Include figure files
\usepackage{dcolumn}% Align table columns on decimal point
\usepackage{bm}% bold math
\usepackage{amsmath}
\usepackage{hyperref}
\usepackage{cleveref}

\UseRawInputEncoding

\begin{document}
\title{Entanglement dynamics of an open moving-biparticle system driven by classical-field}
 \author{Qilin Wang}
\affiliation{Synergetic Innovation Center for Quantum Effects and Application, Key Laboratory of Low-dimensional Quantum Structures and Quantum Control of Ministry of Education, School of Physics and Electronics, Hunan Normal University, Changsha, 410081, People's Republic of China.}
\author{Rongfang Liu}
\affiliation{School of Physical Science and Technology, Lanzhou University, Lanzhou 730000, China.}
\author{Hong-Mei Zou \thanks}
\email{zhmzc1997@hunnu.edu.cn}
\author{Dan Long}
\author{Jia Wang}
 \affiliation{Synergetic Innovation Center for Quantum Effects and Application, Key Laboratory of Low-dimensional Quantum Structures and Quantum Control of Ministry of Education, School of Physics and Electronics, Hunan Normal University, Changsha, 410081, People's Republic of China.}

\date{\today}% It is always \today, today,
             \begin{abstract}
In this work, the entanglement dynamics of a moving-biparticle system driven by an external classical field are investigated, where the moving-biparticle system is coupled with a zero temperature common environment. The analytical expressions of the density operator and the entanglement can be obtained by using the dressed-state basis when the total excitation number is one. We also discuss in detail the effects of different parameters on the entanglement dynamics. The results show that the classical driving can not only protect the entanglement, but also effectively eliminate the influence of the qubit velocity and the detuning on the quantum entanglement.
\end{abstract}

\maketitle

    \section{INTRODUCTION}

 Entanglement is a nonlocal quantum coherence of a composite system, which describes the quantum relationship between two or more quantum subsystems. Quantum entanglement is an important quantum resource in quantum information processing tasks, such as quantum teleportation, quantum dense coding, quantum cryptography, and quantum computing \cite{Amico,Bouwmeester,Plenio,Ekert,Bennett}. Since the quantum entanglement is a pure quantum property, it doesn't dissipate over time only in an ideal isolated system. However, the interaction between any real physical system and its surrounding environment is inevitable, which will lead to the dissipation and the decoherence of quantum systems. Therefore, it is an important subject to study the dynamical evolution of the entangled state caused by the dissipation and the decoherence \cite{Breuer,Duan,Yu,Simon,D¨¹r}.

Numerous investigations on the entanglement dynamics in open quantum systems have been done over the past two decades \cite{Yu 2004,Bellomo,Cao,Maniscalco,Tong,Tan,Jing,Barchielli,Jing 2012,Yang,Franco,Cheng,Aolita,Franco 2015,Mu,Franco 2016,Breuer 2016,Costa-Filho,Leggio,Mortezapour1,Vega,Man}. For examples, Hong-Mei Zou and Mao-Fa Fang studied analytical solution and entanglement swapping of a double Jaynes-Cummings model in non-Markovian environments \cite{Zou}. Wei Jiang $et\ al.$ investigated Non-Markovian entanglement dynamics of open quantum systems with continuous measurement feedback \cite{Jiang}. Bahram Ahansaz $et\ al.$ presented a method of entanglement protection for a two-qutrit V-type open system on the basis of system-reservoir bound states \cite{Ahansaz}. Andrey R. Kolovsky revisited the Born-Markov approximation for an open quantum system by considering a microscopic model of the bath \cite{Kolovsky}. These above results indicate that the memory and feedback of non-Markovian environments have some protective effect on the entanglement dynamics of open quantum systems. However, the memory and feedback effects are usually not sufficient to preserve the quantum entanglement of an open system for long a time. How to maintain entanglement of open quantum systems for a long time is an important challenge in quantum information processing tasks. In past years, some quantum control methods have been proposed in order to effectively protect quantum entanglement of open systems, such as the quantum Zeno effect \cite{Maniscalco,Liu,Petrosky}, weak measurement \cite{Aharonov,Koashi,Yu Liu}, classical field driving \cite{Xiao,Zhang,Liao}, PT-symmetric operation \cite{Guo} and external magnetic field and Dzyaloshinskii-Moriya interaction \cite{Shi-Yang Zhang}. On the other hand, based on the fact that atoms cannot be cooled to a complete standstill in the current cavity QED experiments \cite{Shuo Zhang1,Shuo Zhang2,Shuo Zhang3}, the interaction of moving qubits with electromagnetic radiation \cite{Moustos,Park} and quantum effects of moving-qubit systems \cite{Mortezapour1,Moustos,Park} have been attracting more and more attention in recent years.

In particular, the authors at \cite{Golkar} first investigated the dynamics of entanglement of two moving qubits in a common environment for both strong and weak coupling regimes, and then generalized to the case that an arbitrary number of qubits interact with an environment. They obtained the stationary state of each case in details and also illustrated that how the motion of qubits affects the dynamics of entanglement. The results showed that the movement of the qubits can play a constructive role in protecting of the initial entanglement. Alireza Nourmandipour $et\ al.$ investigated the effect of classical driving on the generation (and protection) of entanglement between two qubits in a cavity. The results show that the classical driving field has a constructive role in the entanglement protection, and show that the classical driving field can protect the entanglement from environmental attenuation\cite{Nourmandipour}. The research results showed that the classical driving can protect the entanglement dynamics of rest qubit systems \cite{Nourmandipour,J. S. Zhang1,Y. J. Zhang,J. S. Zhang2,J. S. Zhang3,Ali Mortezapour}. These works inspire us to investigate the effect of external classical field on the entanglement dynamics of an open moving-biparticle system. Our purpose is to understand whether the classical driving can also protect the entanglement of moving-qubit systems, and we also hope that the influence of the motion of qubits on entanglement can be suppressed through the regulation of classical driving. We found that the classical driving can not only protect the entanglement of moving-qubit systems, but also effectively eliminate the influence of the qubit velocity on the entanglement, which can provide some references in the theoretical and experimental research of open quantum systems.

In this paper, we consider a moving-biparticle system coupled with a zero temperature common environment, where the qubits are driven by an external classical-field. We obtain the analytical expression of the density operator and the entanglement by using the dressed-state basis. We also discuss in detail the effects of different parameters on the entanglement dynamics. The results show that the classical driving can effectively eliminate the influence of the qubit velocity and the detuning on the quantum entanglement.

The paper is organized as follows. The physical model and the analytical solutions are given in the second section. In the third section, we study the entanglement dynamics in resonance and detuning cases, respectively. In the fourth section, we provide a brief summary of this paper.

    \section{PHYSICAL MODEL AND ANALYTICAL SOLUTION}
Based on the Hamiltonian about the moving-qubit in \cite{Golkar} and the Hamiltonian about the classical driving in \cite{Nourmandipour}, we construct a moving-biparticle model driven by classical field, in which the moving-biparticle system couples with a length of L (L approaches infinity), and the particles move along the z-axis and are driven by an external classical-field (see FIG.1). Meanwhile, in order to effectively control the quantum effect of qubits, the classical field is polarized along the x-axis and propagates along the y-axis. At this time, the Hamiltonian of the system is ($\hbar$=1).
\begin{equation}\label{EB201}
\begin{split}
\hat{H} &=\frac{1}{2}\omega_{0}\sum_{i=1}^{2}\hat\sigma_{z}^{i}+\sum_{k}\omega_{k}\hat a_{k}^{+} \hat a_{k}\\
&+\sum_{j=1}^{2} \Omega\left(e^{-i w_{L}t}\hat\sigma_{+}^{j} +e^{i w_{L} t} \hat\sigma_{-}^{j}\right) \\
&+\sum_{j=1}^{2} \sum_{k}\left(f_{k}\left(z_{j}\right) \alpha_{j} \hat\sigma_{+}^{j} g_{k} \hat a_{k}+f_{k}\left(z_{j}\right) \alpha_{j} \hat \sigma_{-}^{j} g_{k}^{*}\hat a_{k}^{+}\right).
\end{split}
\end{equation}%
 $\omega_{0}$ is the transition frequency of the qubits. $\omega_{L}$ and $\omega_{k}$ represent the frequencies of the classical driving field and the cavity quantized modes, respectively. The operators $\hat \sigma_{z}$ and $\hat \sigma_{\pm}$ are defined by $\hat \sigma_{z}=|e\rangle\langle e|-| g\rangle\langle g|$, $\hat \sigma_{+}= |e\rangle\langle g|$, and $\hat \sigma_{-}= |g\rangle\langle e|$ associated with the upper level $|e\rangle$ and lower level $| g\rangle$. $\hat a_{k}^{+}\left(\hat a_{k}\right)$ are the creation(annihilation) operator. In this paper, the dimensionless constant $\alpha_{j}(j=1,2)$ describes the coupling strength between the qubit and the environment, which depends on the relative position of the two qubits in the environment. In addition, $g_{k}$ denotes the coupling constant between the qubits and the k-th mode, $\Omega$ is the classical driving strength of the qubits.
\begin{figure}[h]
\includegraphics[width=8cm,height=5cm]{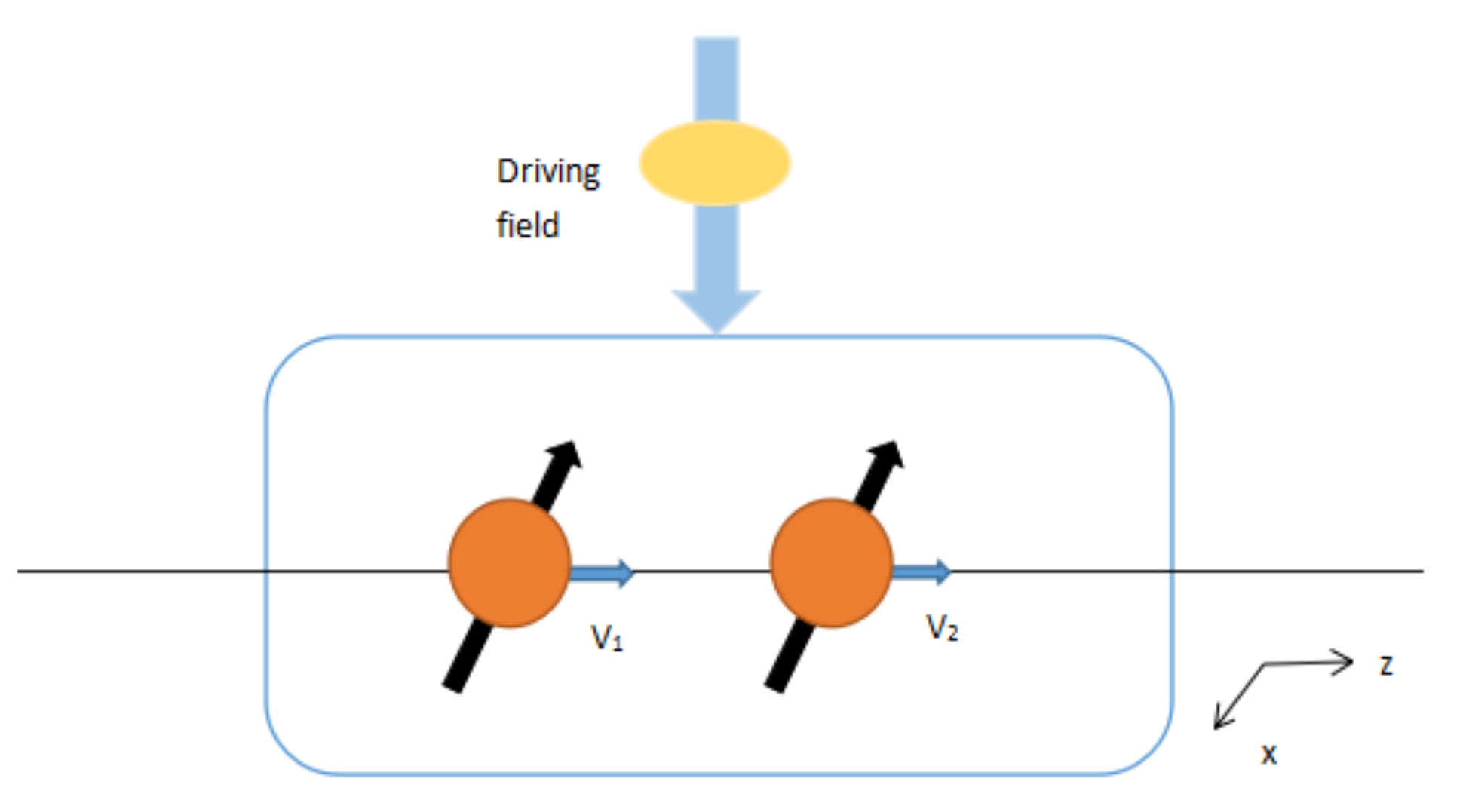}
\caption{Schematic illustration of a setup where two qubits are moving inside a cavity and driven by the classical field. The qubits are two-level atoms with transition frequency $\omega_{0}$ travelling with constant velocity $v_{j}$.}
\label{fig:1}
\end{figure}
When the qubits move along the z-axis, the dependency function $f_{k}(z_{j})$ \cite{Leonardi} induced by the velocity of the qubits motion can be expressed as
\begin{equation}\label{EB202}
f_{k}(z_{j})=f_{k}(v_{j}t)=\sin [k(z_{j}-l)]=\sin [\omega_{k}(\beta_{j} t-\tau)].
\end{equation}%
where $\beta_{j}=v_{j}/c$ and $\tau =L/c$ with $L$ being the size of the cavity. $v_{j}(j=1,2)$ represents the velocity of the i-th qubit, $c$ means the speed of light. Because the unitary transformation does not change its eigenvalues, we do a rotation transformation on the Hamiltonian in Eq.(1), i.e. $U_{R}=\exp{[-i\omega_{L}(\hat \sigma_{z}^{1}+\hat \sigma_{z}^{2})t/2]}$, and the Hamiltonian becomes is
\begin{equation}\label{EB203}
\begin{split}
\hat{H}_{e} &=\frac{\Delta}{2} (\hat\sigma_{z}^{1}+\hat\sigma_{z}^{2})+\Omega (\hat\sigma_{x}^{1}+\hat\sigma_{x}^{2})+\sum_{k} \omega_{k} \hat a_{k}^{+} \hat a_{k}\\
&+\sum_{k} \left[g_{k} \hat a_{k}f_{k}(z_{1})\hat\sigma _{+}^{1}\alpha _{1} e^{i \omega_{L} t} +\text {h.c. }\right]\\
&+\sum_{k} \left[g_{k}\hat a_{k}f_{k}(z_{2})\hat \sigma _{+}^{2}\alpha _{2} e^{i \omega_{L} t} +\text {h.c. }\right],
\end{split}
\end{equation}%
where $\Delta=\omega_{0}-\omega_{L}$, we diagonalize the first two items on the right side of the above formula.
\begin{equation}\label{EB204}
\hat{H}_{e}^{1}=\frac{\omega_{D}}{2} (\rho_{z}^{1}+\rho_{z}^{2}),
\end{equation}%
in which $\omega_{D}$ and $\rho_{z}$ are the transition frequency and the inversion operators in the dressed-state basis.
\begin{equation}\label{EB205}
\omega_{D}=\sqrt{\Delta^{2}+4|\Omega|^{2}},
\end{equation}%
\begin{equation}\label{EB206}
\rho_{z}=|E\rangle\langle E|-| G\rangle\langle G|,
\end{equation}%
$|E\rangle$ and $| G\rangle$ are the excited state and the ground state respectively under the dressed-state representation.
\begin{equation}\label{EB207}
|E\rangle=\cos\frac{\theta }{2}|e\rangle+\sin \frac{\theta }{2}|g\rangle,
\end{equation}%
\begin{equation}\label{EB208}
|G\rangle=\cos\frac{\theta }{2}|g\rangle-\sin \frac{\theta }{2}|e\rangle,
\end{equation}%
In the above relation $\theta =\tan^{-1}(\frac{2\left |\Omega \right|}{\Delta })$, in the dressed-state representative, the effective Hamiltonian can be expressed as
\begin{equation}\label{EB209}
\begin{split}
\hat{H}^{D}_{e} &=\frac{\omega_{D}}{2} (\rho_{z}^{1}+\rho_{z}^{2})+\sum_{k} \omega_{k} \hat a_{k}^{\dagger} \hat a_{k}\\
&+\cos^{2} \frac{\theta }{2} \sum_{k} \left[g_{k}\hat a_{k}f_{k}(z_{1})\rho_{+}^{1}\alpha _{1} e^{i \omega_{L} t} +\text {h.c. }\right]\\
&+\cos^{2} \frac{\theta }{2} \sum_{k} \left[g_{k}\hat a_{k}f_{k}(z_{2})\rho_{+}^{2}\alpha _{2} e^{i \omega_{L} t} +\text {h.c. }\right],
\end{split}
\end{equation}%
and ascending operator $\rho_{+}$ is defined as $\rho_{+}=|E\rangle\langle G|$. Under the rotating transformation
$U_{I}=\exp{[-i\omega_{D}(\rho_{z}^{1}+\rho_{z}^{2})t/2+\omega_{k}\hat a_{k}^{+} \hat a_{k} t]}$, we can obtain the interaction Hamiltonian as
\begin{equation}\label{EB210}
\begin{split}
	\hat{H}_{I}&= \cos^{2}\frac{\theta }{2} \sum_{k}\sum_{j=1}^{2}\left[f_{k}(z_{j})g_{k} \hat a_{k}\alpha _{j} \rho_{+}^{j} e^{i\left(\omega_{D}+\omega_{L}-\omega_{k}\right) t}\right.\\
	&\left.+f_{k}(z_{j})g_{k}^{*}\hat a_{k}^{\dagger}\alpha _{j}\rho_{-}^{j} e^{-i\left(\omega_{D}+\omega_{L}-\omega_{k}\right) t}\right].
\end{split}
\end{equation}%

Assume that the total system is initially in
\begin{equation}\label{EB211}
\left|\psi{(0)}\right\rangle=\left(\cos \frac{\eta }{2}|E,G\rangle+\sin \frac{\eta }{2}e^{i\phi }|G,E\rangle\right)\left |0\ \right \rangle_{R}.
\end{equation}%
in which $\left |0\ \right \rangle_{R}$ is the multi-mode vacuum state of environment. Supposing that the total excitation number is 1, the time evolution state of the system is given by
\begin{equation}\label{EB212}
\begin{split}
|\psi(t)\rangle &=c_{1}(t)|E,G\rangle\left |0\ \right \rangle_{R}+c_{2}(t)|G, E\rangle\left |0\ \right \rangle_{R}\\
&+\sum c_{k}(t)|G, G\rangle\left|1_{k}\right\rangle.
\end{split}
\end{equation}%
with $|1_{k}\rangle$ being the state of the environment with only one excitation in the k-th mode. By solving the $Schr\ddot{o}dinger$ equation
\begin{equation}\label{EB213}
i\hbar\frac{d}{d t}|\psi(t)\rangle_{I}=H_{I}|\psi(t)\rangle.
\end{equation}%
the differential equations of $c_{1}(t)$ and $c_{2}(t)$ and $c_{k}$ are expressed as
\begin{equation}\label{EB214}
\begin{split}
\dot{c}_{1}(t)&=-i\cos^{2} \frac{\theta }{2}\sum_{k} g_{k}f_{k}(z_{1})\alpha_{1}c_{k}(t) e^{i\left(\omega_{D}+\omega_{L}-\omega_{k}\right) t},\\
\dot{c}_{2}(t)&=-i\cos^{2} \frac{\theta }{2}\sum_{k} g_{k}f_{k}(z_{2})\alpha_{2}c_{k}(t) e^{i\left(\omega_{D}+\omega_{L}-\omega_{k}\right)t} ,
\end{split}
\end{equation}%
\begin{equation}\label{EB215}
\begin{split}
\dot{c}_{k}(t)&=-i\cos^{2}\frac{\theta}{2}g_{k}^{*}e^{-i\left(\omega_{D}+\omega_{L}-\omega_{k}\right)t}  \\
               &[f_{k}(z_{1})\alpha_{1}c_{1}(t)+f_{k}(z_{2})\alpha_{2}c_{2}(t)].
\end{split}
\end{equation}%

Let $v_{1}=v_{2}=v$, solving Eq.~(\ref{EB215}) and then substituting into Eq.~(\ref{EB214}), we get differential equations about $c_{1}(t)$ and $c_{2}(t)$
\begin{equation}\label{EB216}
\begin{split}
\dot{c}_{1}(t)&=-\cos^{4} \frac{\theta }{2}\int_{0}^{t} d t^{\prime } F\left(t,t^{\prime }\right)[\alpha^{2}_{1}c_{1}(t^{\prime })+\alpha _{1}\alpha _{2}c_{2}(t^{\prime })],\\
\dot{c}_{2}(t)&=-\cos^{4} \frac{\theta }{2}\int_{0}^{t} d t^{\prime } F\left(t,t^{\prime }\right)[\alpha^{2}_{2}c_{2}(t^{\prime })+\alpha _{1}\alpha _{2}c_{1}(t^{\prime })],
\end{split}
\end{equation}%
and $ F\left(t,t^{\prime }\right)$ is the correlation function of the environment.
\begin{equation}\label{EB217}
\begin{split}
	F\left(t,t^{\prime }\right)&= \int_{0}^{\infty} J\left(\omega_{k}\right) \sin \left[\omega_{k}\left(\beta t-\tau\right)\right] \times \\
	& \sin \left[\omega_{k}\left(\beta t^{\prime }-\tau\right)\right] e^{i\left(\omega_{D}+\omega_{L}-\omega_{k}\right)\left(t-t^{\prime }\right)} d \omega_{k}.
\end{split}
\end{equation}%
 $J\left(\omega_{k}\right)$ is the spectral density with Lorentz form i.e
\begin{equation}\label{EB218}
J\left(\omega_{k}\right)=\frac{1}{2 \pi} \frac{\gamma \lambda^{2}}{(\omega_{0}-\omega _{k})^{2}+\lambda ^{2}}.
\end{equation}%
 here the center frequency of the spectrum is equal to $\omega_{0}$. $\lambda $ is the spectral width of the environment, and $\gamma$ is the coupling strength. $\lambda > 2\gamma $ indicates that the qubit-environment is weak-coupling, $\lambda < 2\gamma$ means that the qubit-environment is strong-coupling \cite{Breuer,Heinz-Peter Breuer,Hong-Mei Zou}. Substituting the Eq.~(\ref{EB218}) into the Eq.~(\ref{EB217}) the following expression can be obtained
\begin{equation}\label{EB219}
F\left(t,t^{\prime }\right)=\sum_{i=1}^{4} F_{i}\left(t,t^{\prime }\right).
\end{equation}%

Making use of
\begin{equation}\label{EB220}
\int_{-\infty}^{\infty} \frac{e^{-iza }}{z^{2}+\lambda^{2}} d z=\frac{\pi}{\lambda} e^{-\lambda|a|}.
\end{equation}%
and $0< \beta < 1$, then we have
\begin{equation}\label{EB221}
\begin{split}
F_{1}\left(t, t^{\prime}\right)&=\frac{-r \lambda}{8} e^{iw_{0}\beta \left(t+t^{\prime}\right)-2i w_{0}\tau+i(w_{D}-\Delta )(t-t^{\prime })} \\
&\times e^{-\lambda[-\beta(t+t^{\prime})+2 \tau+(t-t^{\prime})]},
\end{split}
\end{equation}%
\begin{equation}\label{EB222}
F_{2}\left(t, t^{\prime}\right)=\frac{r \lambda}{8} e^{i(w_{D}-\Delta+w_{0}\beta )(t-t^{\prime })} e^{-\lambda[(1-\beta)(t-t^{\prime})]},
\end{equation}%
\begin{equation}\label{EB223}
F_{3}\left(t, t^{\prime}\right)=\frac{r \lambda}{8} e^{i(w_{D}-\Delta-w_{0}\beta )(t-t^{\prime })} e^{-\lambda[(1+\beta)(t-t^{\prime})]},
\end{equation}%
\begin{equation}\label{EB224}
\begin{split}
F_{4}\left(t, t^{\prime}\right)&=\frac{-r \lambda}{8} e^{-iw_{0}\beta \left(t+t^{\prime}\right)+2i w_{0}\tau+i(w_{D}-\Delta )(t-t^{\prime })} \\
&\times e^{-\lambda[-\beta(t+t^{\prime})+2 \tau-(t-t^{\prime})]},
\end{split}
\end{equation}%

 In this work, we consider the case that a length of the cavity is infinity, i.e., $\tau $ $\to$ $\infty$, Eq.~(\ref{EB219}) is simplified as
\begin{equation}\label{EB225}
\begin{split}
F\left(t, t^{\prime}\right)&=\frac{r \lambda}{8} e^{\left\{\left(\lambda+i \omega_{0}\right) \beta-\left[\lambda+i\left(\Delta-\omega_{D}\right)\right]\right\}(t-t^{\prime})}\\
& +\frac{r \lambda}{8} e^{\left\{-\left(\lambda+i \omega_{0}\right) \beta-\left[\lambda+i\left(\Delta-\omega_{D}\right)\right]\right\}(t-t^{\prime})}\\
&=\frac{r \lambda}{8}\left[e^{V_{+}(t-t^{\prime })}+e^{V_{-}\left(t-t^{\prime}\right)}\right].
\end{split}
\end{equation}%
with $y=\lambda +i(\Delta -\omega _{D})$, $\mu =\lambda +i\omega _{0}$, $V_{+}=-y+\mu \beta $, $V_{-}=-y-\mu \beta$.

Then, we bring Eq.~(\ref{EB225}) into Eq.~(\ref{EB216}), we can obtain the exact solution directly by the Laplace approach as
\begin{equation}\label{EB226}
\begin{split}
c_{1}(t)&=(r_{2}^2\cos\frac{\eta}{2}-r_{1}r_{2}\sin \frac{\eta}{2}e^{i\phi })+\\
&(r_{1}^{2}\cos\frac{\eta}{2} +r_{1}r_{2}\sin \frac{\eta}{2}e^{i\phi})\varepsilon (t),
\end{split}
\end{equation}%
\begin{equation}\label{EB227}
\begin{split}
c_{2}(t)&=(r_{1}^2\sin \frac{\eta }{2}e^{i\phi } -r_{1}r_{2}\cos \frac{\eta }{2})+\\
&(r_{2}^{2}\sin \frac{\eta }{2}e^{i\phi } +r_{1}r_{2}\cos \frac{\eta }{2})\varepsilon (t),
\end{split}
\end{equation}%
where $\alpha_{T}=\left(\alpha_{1}^{2}+\alpha_{2}^{2}\right)^{1 / 2}$, $r_{1}=\alpha_{1}/ \alpha_{T}$ and $r_{2}=\alpha_{2}/ \alpha_{T}$. To simplify the expression, we introduce the set coupling constant $\alpha_{T}$ and make $\alpha_{T}$=1. The relative coupling constant $r_{j}$ (j=1,2) and $r_{1}^{2}$+$r_{2}^{2}$=1, so $r_{1}$ is considered as an independent variable.
\begin{equation}\label{EB228}
\begin{split}
\varepsilon (t)&=\frac{\left(q_{1}-y_{+}\right)\left(q_{1}-y_{-}\right)}{\left(q_{1}-q_{2}\right)\left(q_{1}-q_{3}\right)}e^{q_{1}yt}\\
 &+\frac{\left(q_{2}-y_{+}\right)\left(q_{2}-y_{-}\right)}{\left(q_{2}-q_{1}\right)\left(q_{2}-q_{3}\right)} e^{q_{2}yt}\\
 &+\frac{\left(q_{3}-y_{+}\right)\left(q_{3}-y_{-}\right)}{\left(q_{3}-q_{1}\right)\left(q_{3}-q_{2}\right)} e^{q_{3}yt},
\end{split}
\end{equation}%

\begin{equation}\label{EB229}
q^{3}+2q^{2}+q(y_{+}y_{-}+\frac{\gamma \lambda }{4y^2}\cos^{4}\frac{\theta }{2})+\frac{\gamma \lambda }{4y^2}\cos^{4} \frac{\theta }{2}=0.
\end{equation}%
with $y_{+}=\frac{V_{+}}{y}$, $y_{-}=\frac{V_{-}}{y}$, and $q_{1}$, $q_{2}$, $q_{3}$ are the three solutions of Eq.~(\ref{EB229}).

In the $\{|E\rangle,|G\rangle\}$ basis, the reduced density operator for the biparticles is given by
\begin{equation}\label{EB230}
\begin{split}
\rho(t)&=\left(\begin{array}{cccc}
	0 & 0 & 0 & 0 \\
	0 & \left|c_{1}(t)\right|^{2} & c_{1}(t) c_{2}^{*}(t) & 0 \\
	0 & c_{1}^{*}(t) c_{2}(t) & \left|c_{2}(t)\right|^{2} & 0 \\
	0 & 0 & 0 & 1-\left|c_{1}\right|^{2}-\left|c_{2}\right|^{2}
\end{array}\right).
\end{split}
\end{equation}%

\section{ENTANGLEMENT DYNAMICS}
In this paper, we choose Wootters concurrence \cite{William} to quantify entanglement, which is defined as
\begin{equation}\label{EB301}	
C_{\hat{\rho}}(t)=\max \left\{0, \sqrt{l_{1}}-\sqrt{l_{2}}-\sqrt{l_{3}}-\sqrt{l_{4}}\right\},
\end{equation}%
where $l_{i}$ is the eigenvalue of matrix $\tilde{\rho}^{\Phi}$ in decreasing order,
\begin{equation}\label{EB302}
\tilde{\rho}^{\Phi}=\hat{\rho}(t)\left(\hat{\sigma}_{y} \otimes \hat{\sigma}_{y}\right) \hat{\rho}(t)^{*}\left(\hat{\sigma}_{y} \otimes \hat{\sigma}_{y}\right),
\end{equation}%
It is well-known that $\hat{\sigma_{y}}$ represents Pauli matrices and $\hat{\rho}(t)^{*}$ is the complex conjugation of $\hat{\rho}(t)$. $C_{\hat{\rho}}(t)$ ranges from 0 to 1, representing the disentangled state to the maximally entangled state. For the density matrix given in Eq.~(\ref{EB230}), we can know that the entanglement is
\begin{equation}\label{EB303}
C_{\hat{\rho}}(t)=2\left|c_{1}(t) c_{2}^{*}(t)\right|.
\end{equation}%

 \subsection{Entanglement dynamics in resonance case}
 Now we discuss the entanglement dynamics of the moving-biparticle system driven by an external classical field when $\Delta =w_{0}-w_{L}=0$. We consider respectively the entanglement dynamics in the weak and strong coupling regimes, i.e, for $\lambda > 2\gamma$ and $\lambda <2\gamma $.
\begin{figure}[h]
\includegraphics[width=3.8cm,height=3.5cm]{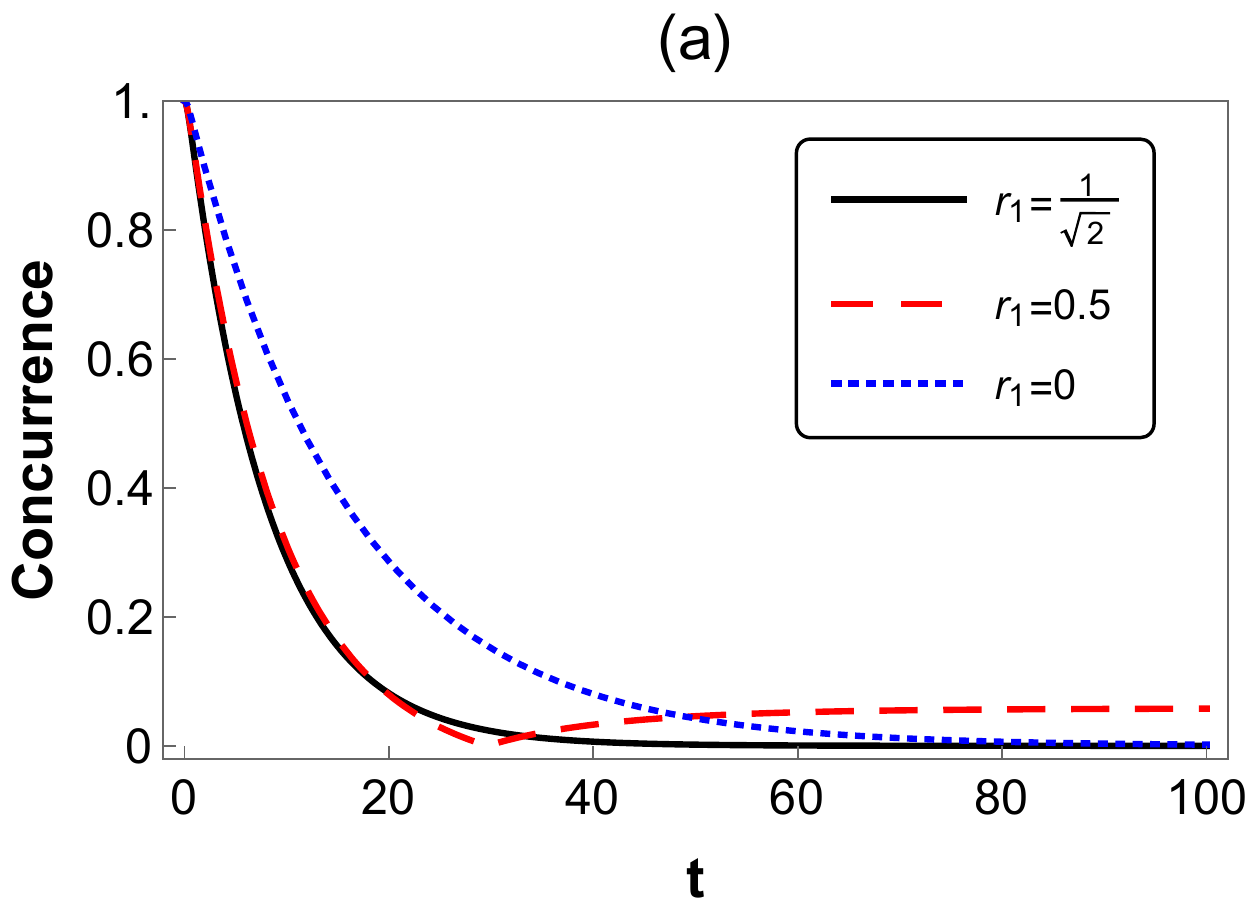}
\includegraphics[width=3.8cm,height=3.5cm]{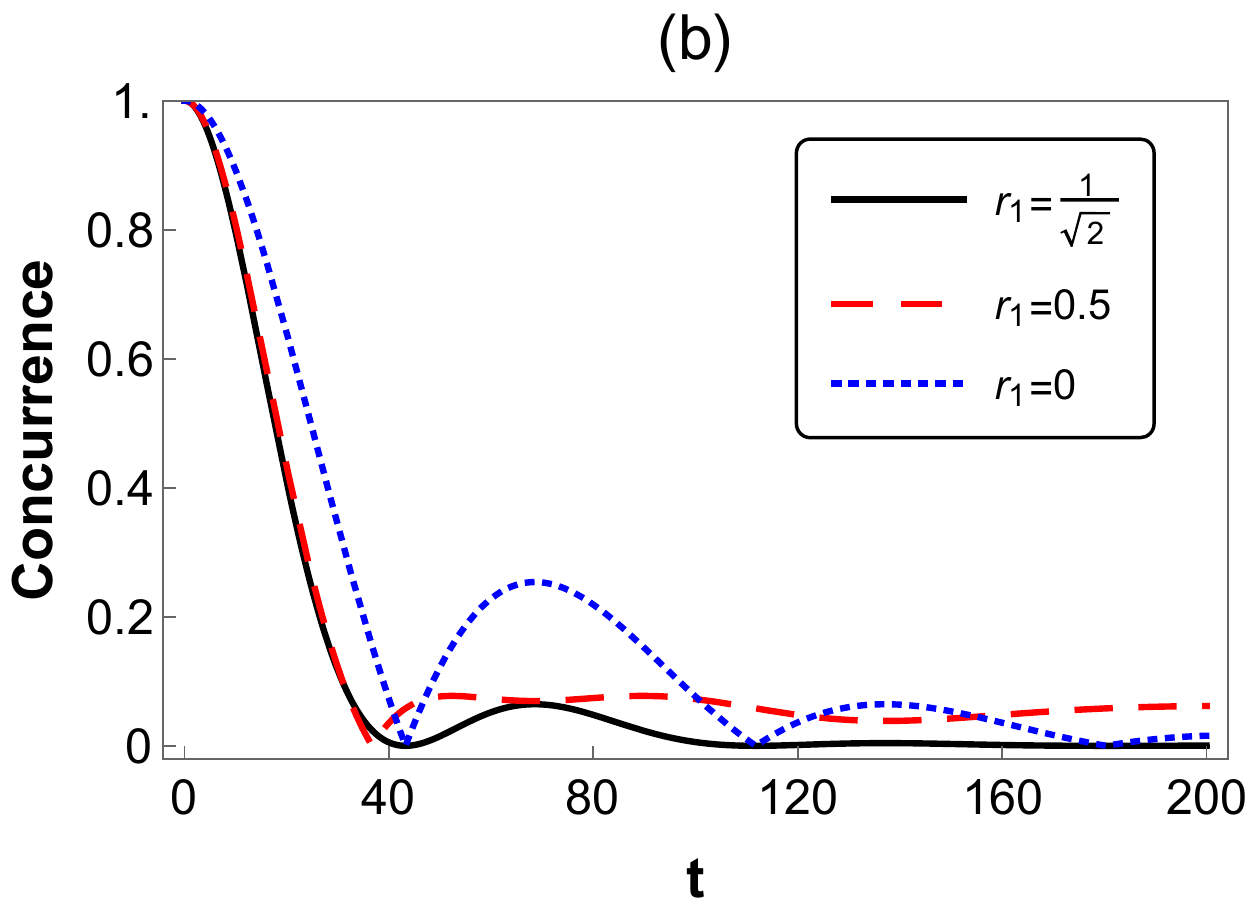}
\caption{Concurrence as a function time t for different $r_{1}$. $r_{1}$=$\frac{1}{\sqrt{2}}$ (the black solid line), $r_{1}$=0.5 (the red dashing line) and $r_{1}$ =0 (the blue dotted line). (a) shows the entanglement evolution under the weak coupling regime ($\lambda=4\gamma$), and (b) shows the entanglement evolution under the strong coupling regime ($\lambda =0.04\gamma$). Other parameters are $\omega _{0}=1.5 \times 10^{9}$, $\gamma =1$, $\beta =0$, $\Omega =0$, $\phi=0$, and $\eta =\frac{\pi }{2}$.}
\label{fig:2}
\end{figure}

  Fig. 2 shows the concurrence as a function of t without the velocity ($\beta =0$) and the driving field ($\Omega =0$) in the weak or strong coupling regimes, respectively. For three different values of the coupling parameter $r_{1}$, namely $r_{1}$=$\frac{1}{\sqrt{2}}$, 0.5, 0. $r_{1}$= $\frac{1}{\sqrt{2}}$ corresponds to the same coupling strength between the two qubits and the environment, and $r_{1}$=0.5 corresponds to the stationary entanglement \cite{Maniscalco,Golkar,Nourmandipour2}. Finally $r_{1}$=0 describes a case in which one of the two qubits is effectively decoupled.

  Fig. 2(a) shows in the weak coupling regime, that the entanglement of the system will all monotonically decay to zero for $r_{1}$ = 0 and $r_{1}$ =$\frac{1}{\sqrt{2}}$. Their difference is that the decay rate of entanglement for $r_{1}$ = 0 is smaller than that for $r_{1}$ =$\frac{1}{\sqrt{2}}$. The reason is that the attenuation of information in the single channel is slower than that in the dual channel in the weak coupling regime. When $r_{1}$=0.5, we find the entanglement of the system will decay with time and then return to a very small stationary value.  Fig. 2(b) shows that, the entanglement dynamics will appear damping oscillation phenomenon due to the memory and feedback effects of non-Markovian environments in the strong coupling regime. And we observed that the entanglement recovery amplitude is the largest when $r_{1}$ = 0.

\begin{figure}[h]
\includegraphics[width=3.8cm,height=3.5cm]{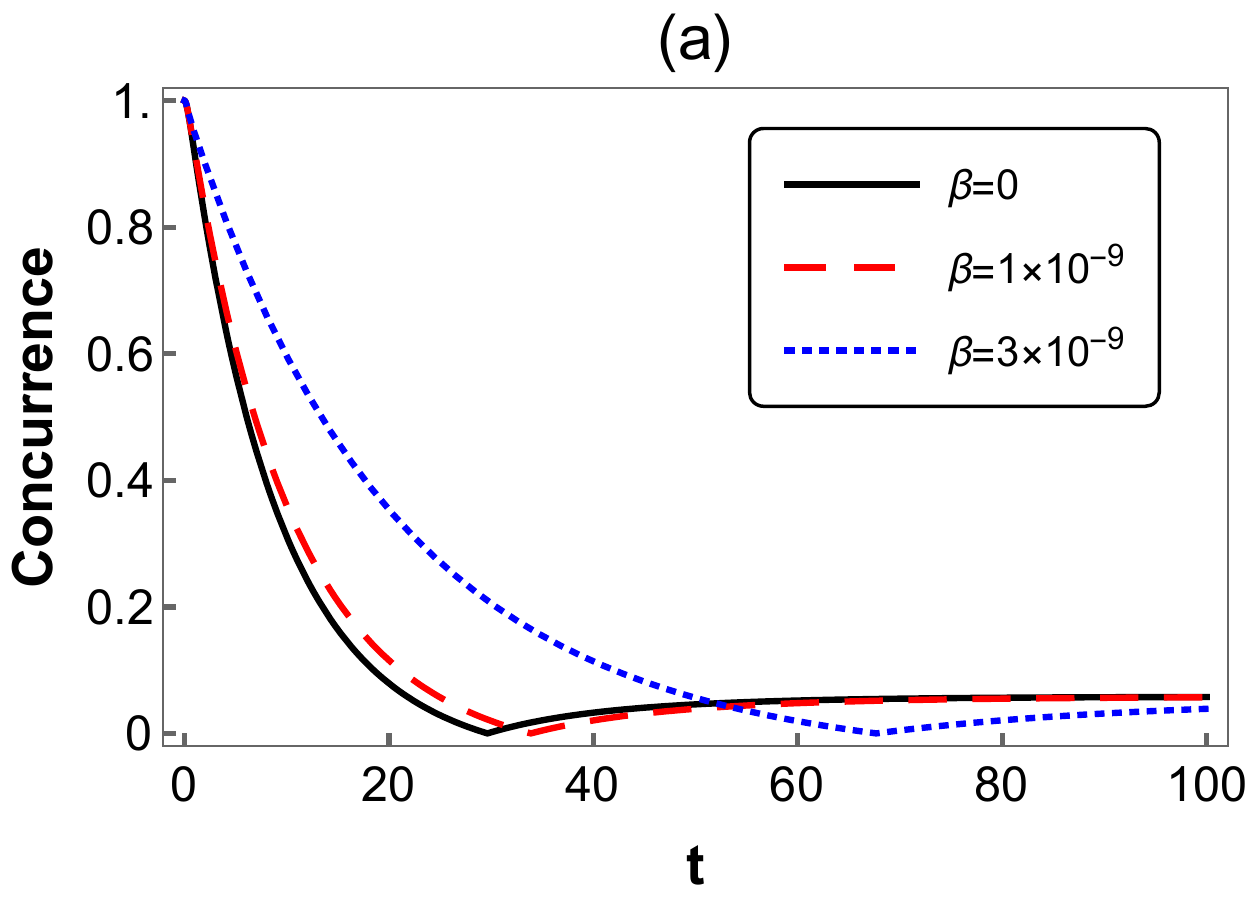}
\includegraphics[width=3.8cm,height=3.5cm]{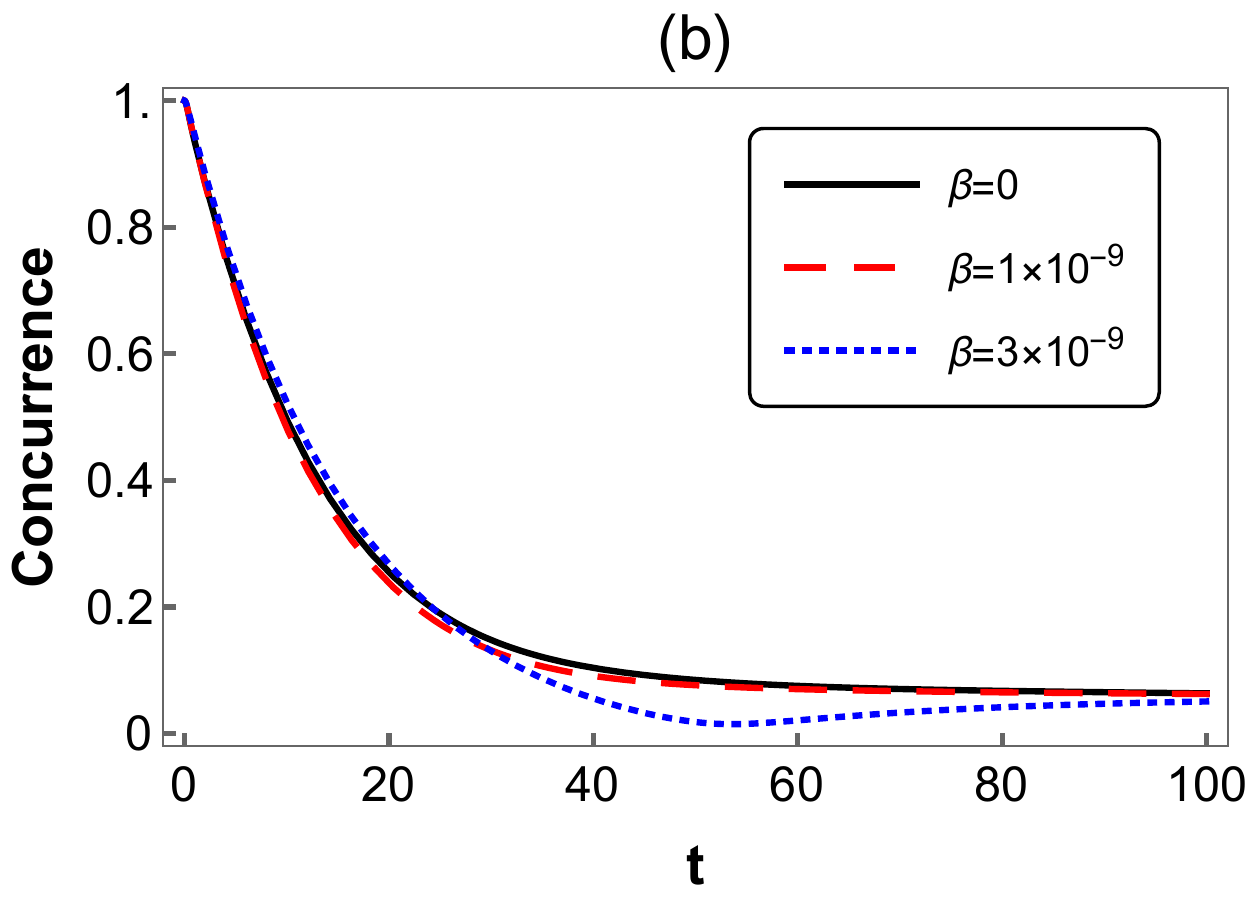}
\caption{Concurrence as a function of time t for different qubits velocity and driving strength in the weak coupling regime($\lambda=4\gamma$). (a) shows the entanglement evolution without driving strength ($\Omega =0$), and (b) shows the entanglement evolution under driving strength ($\Omega =1.6$). Other parameters are $\omega _{0}=1.5 \times 10^{9}$, $\gamma =1$, $\phi=0$, $r_{1}$ =0.5 and $\eta =\frac{\pi }{2}$.}
\label{fig:3}
\end{figure}

   Fig. 3 depicts the concurrence as a function of time t for different driving strength and qubits velocity in the weak coupling regime($\lambda=4\gamma$). Fig. 3(a) shows that without driving strength, the velocity of qubits have an observable influence on the decay rate of entanglement. From this figure, we can see that the entanglement will quickly decay monotonously to zero and then reach the steady value when $\beta =0$. As the velocity of qubits increases, the decay of entanglement will become slower and can survive for a longer time. Fig. 3(b) shows that concurrence as a function of time t for different qubits velocity when $\Omega =1.6$. First, compared to the black solid line in Fig. 3(a) and Fig. 3(b), we can observe that the classical driving slow down the entanglement decay of the static qubit in the weak coupling regime. The results are consistent with that presented in \cite{Nourmandipour}. Then, we can see that the concurrence curves of $\beta =0$ and $\beta =1 \times 10^{-9}$ are close to that of $\beta =3 \times 10^{-9}$, the classical driving can delay the entanglement attenuation and eliminate the influence of the moving velocity on the quantum entanglement. This shows that the driving strength $\Omega $ can regulate the influence of velocity $\beta $ on the entanglement dynamics and can effectively protect entanglement.

\begin{figure}[h]
\includegraphics[width=3.8cm,height=3.5cm]{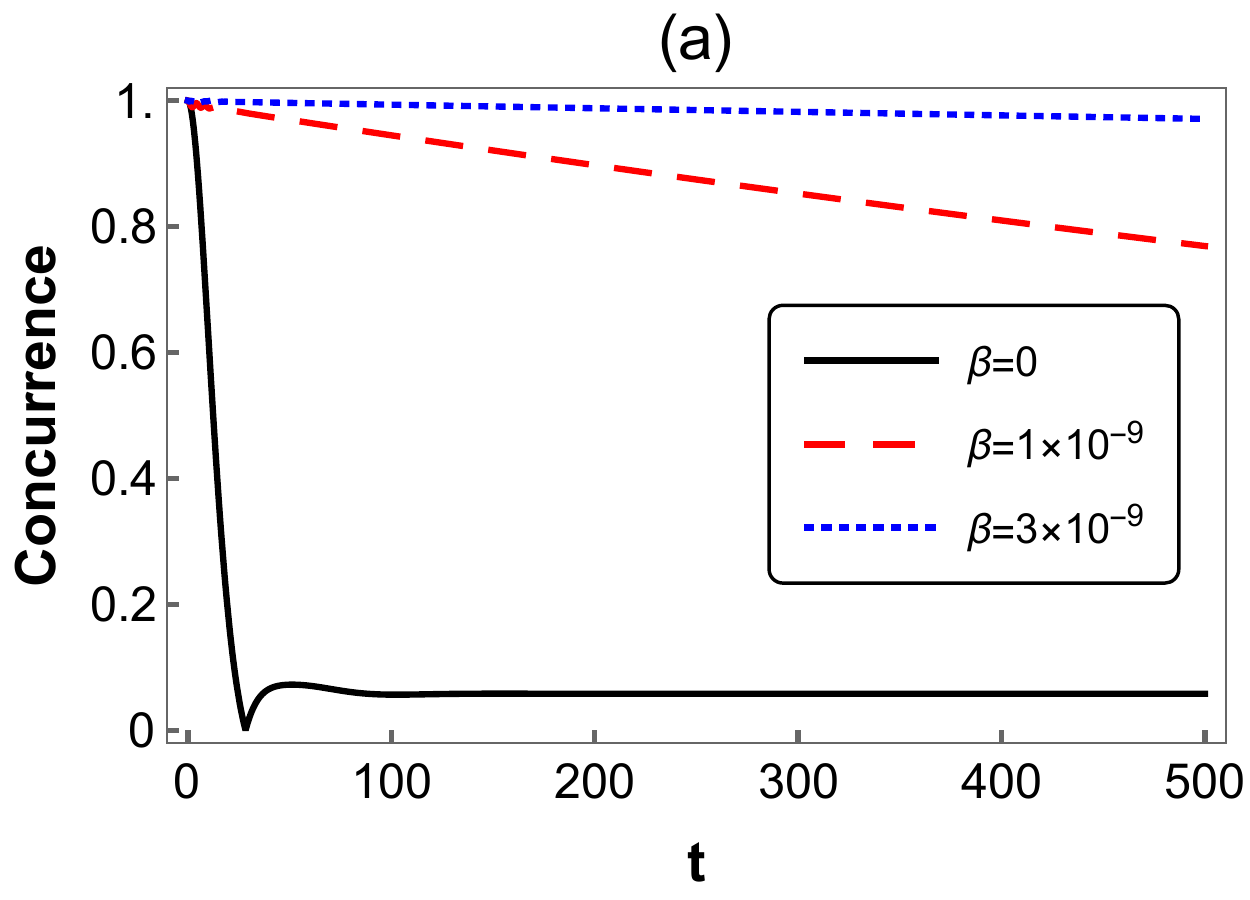}
\includegraphics[width=3.8cm,height=3.5cm]{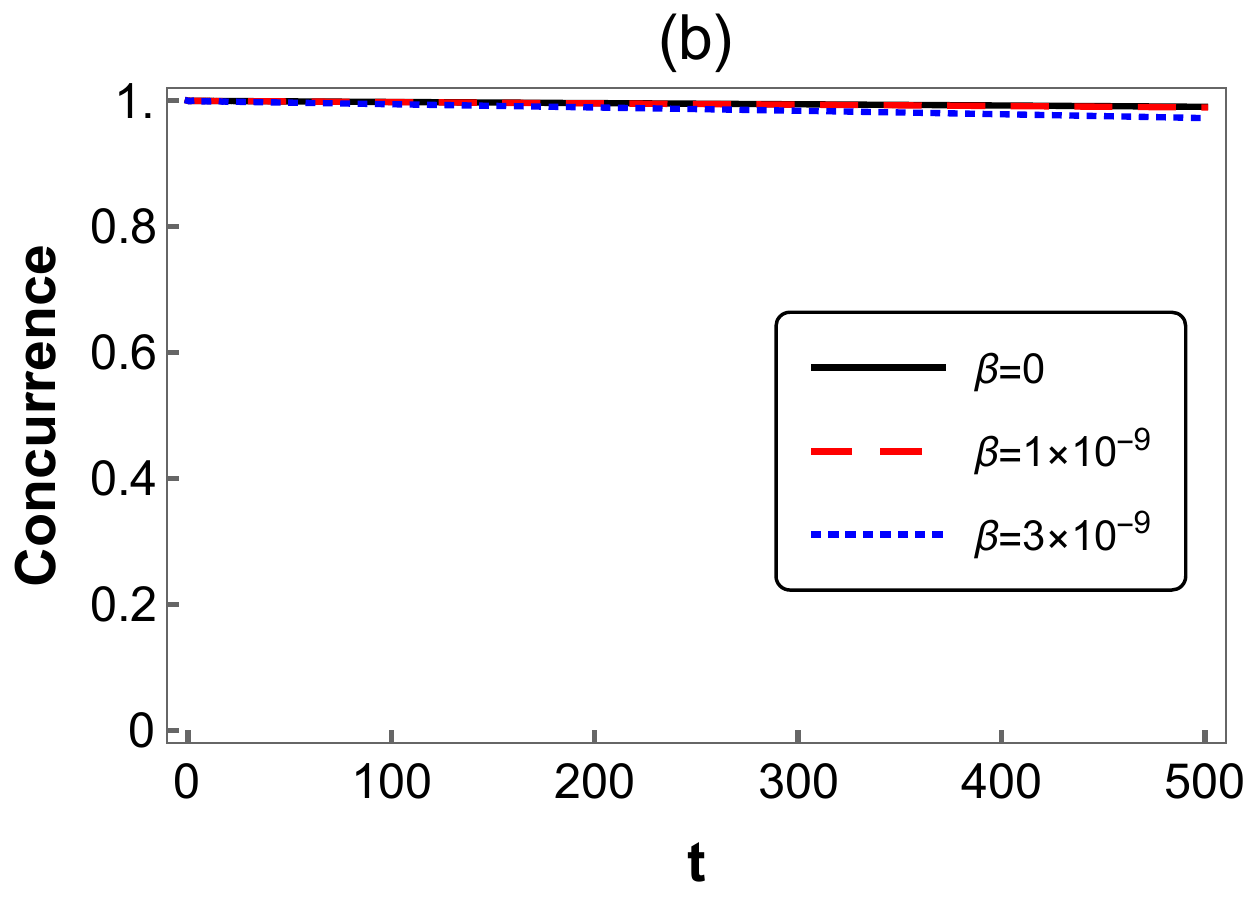}
\caption{Concurrence as a function of time t for different qubits velocity and driving strength in the strong coupling regime ($\lambda=0.1\gamma$). (a) shows the entanglement evolution without driving strength ($\Omega =0$), and (b) shows the entanglement evolution under larger driving strength ($\Omega =4$). Other parameters are $\omega _{0}=1.5 \times 10^{9}$, $\gamma=1$, $\phi=0$, $r_{1}$=0.5 and $\eta=\frac{\pi}{2}$.}
\label{fig:4}
\end{figure}

  In order to understand how the moving-velocity and the driving strength adjust the concurrence in the strong coupling regime, we give Fig. 4. From Fig. 4(a), we know that, when $\beta =0$, the entanglement will quickly decay to zero and then return the very steady value. As the moving-velocity increases, the decay of entanglement will become very slow. The entanglement of the initial state will be well protected due to the memory and feedback effects of non-Markovian environments in the strong coupling regime. The motion of qubits has a significant influence on the survival of entanglement, which is similar to the case of the weak coupling regime. Fig. 4(b) displays the of entanglement dynamic curves when the driving strength $\Omega =4$. Compared with Fig. 4(a), we find that the proper driving strength can well protect the initial entanglement of systems with different velocity. In particular, the protective effect of classical driving of entanglement is the most obvious for the static system ($\beta=0$). Under the classical field driving, the sudden death of the entangled state of the static qubit apparently disappears, as shown the black solid lines in Fig. 4(b), These are also consistent with the results presented in\cite{Nourmandipour}. If there is not the classical driving, the decay rate of entanglement is significantly different from the quantum systems with different velocities. But, under the classical driving, they can almost always keep their initial entanglement, as shown the three coincident curves in Fig. 4(b). That is to say, the classical driving can not only effectively protect the entanglement, but also eliminate the influence of the particle velocity on entanglement.

  Therefore, the proper qubit-environment coupling $r_{1}$ and velocity ratio $\beta$, the smaller spectral width $\lambda$ and the stronger classical driving $\Omega$ can all effectively protect the quantum entanglement of the moving-biparticle system. Especially, in the strong coupling regime, the classical driving can not only effectively protect entanglement, but also eliminate the influence of the particle velocity on entanglement.

 \subsection{Entanglement dynamics in detuning case}
 Nextly, we consider the entanglement dynamics of the moving-biparticle system driven by an external classical-field when $\Delta=\omega _{0}-\omega _{L}\ne0$ (i.e., in the detuning case)

\begin{figure}[h]
\includegraphics[width=3.8cm,height=3.5cm]{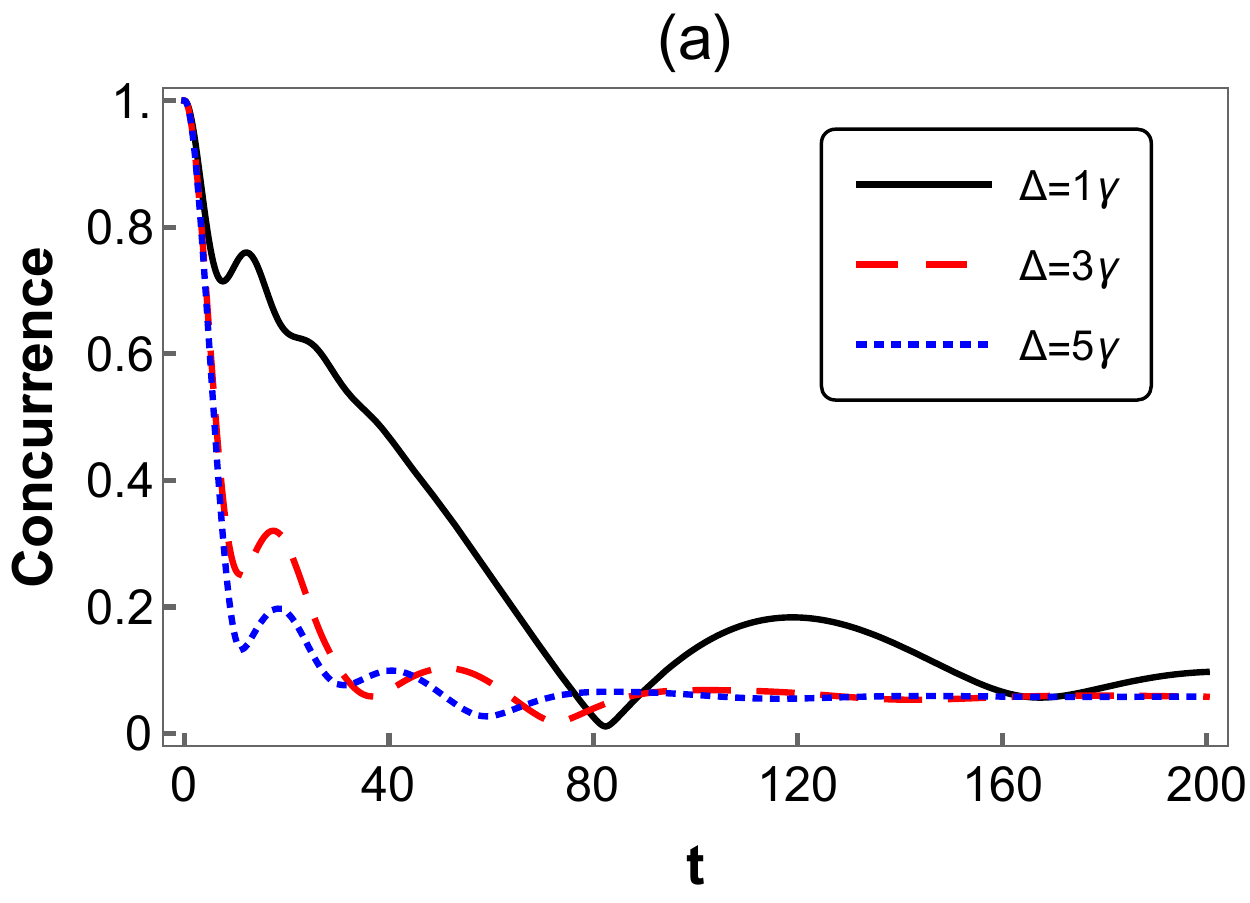}
\includegraphics[width=3.8cm,height=3.5cm]{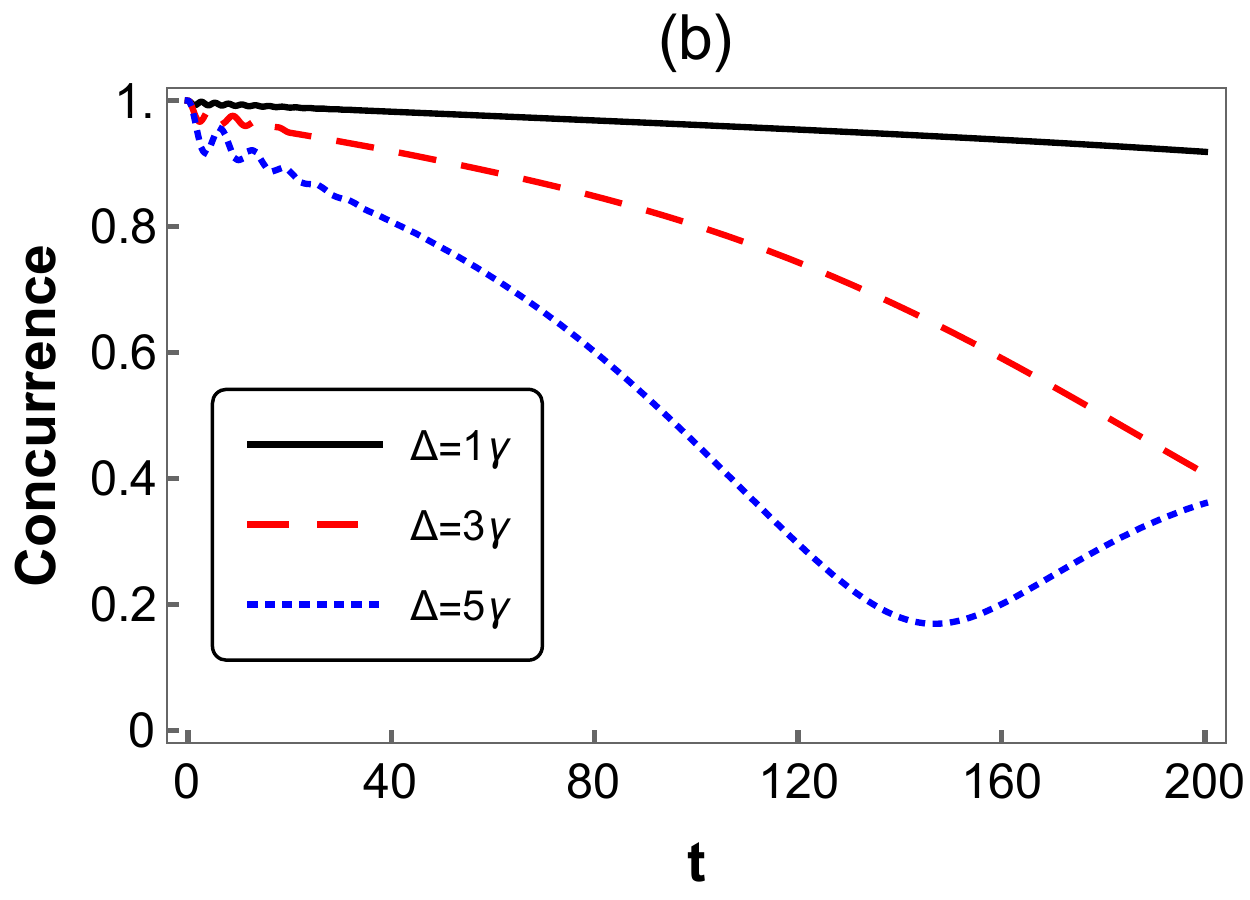}
\includegraphics[width=3.8cm,height=3.5cm]{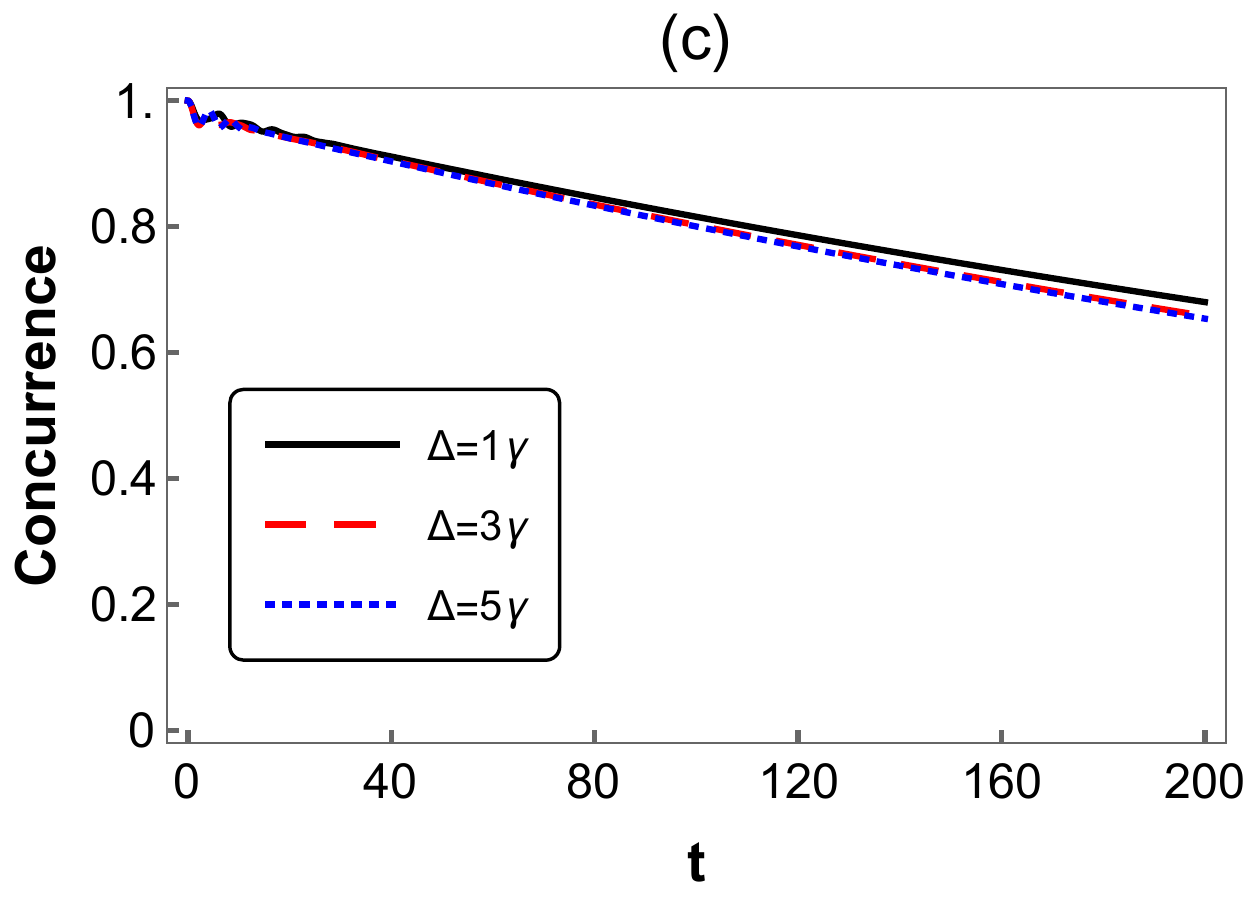}
\includegraphics[width=3.8cm,height=3.5cm]{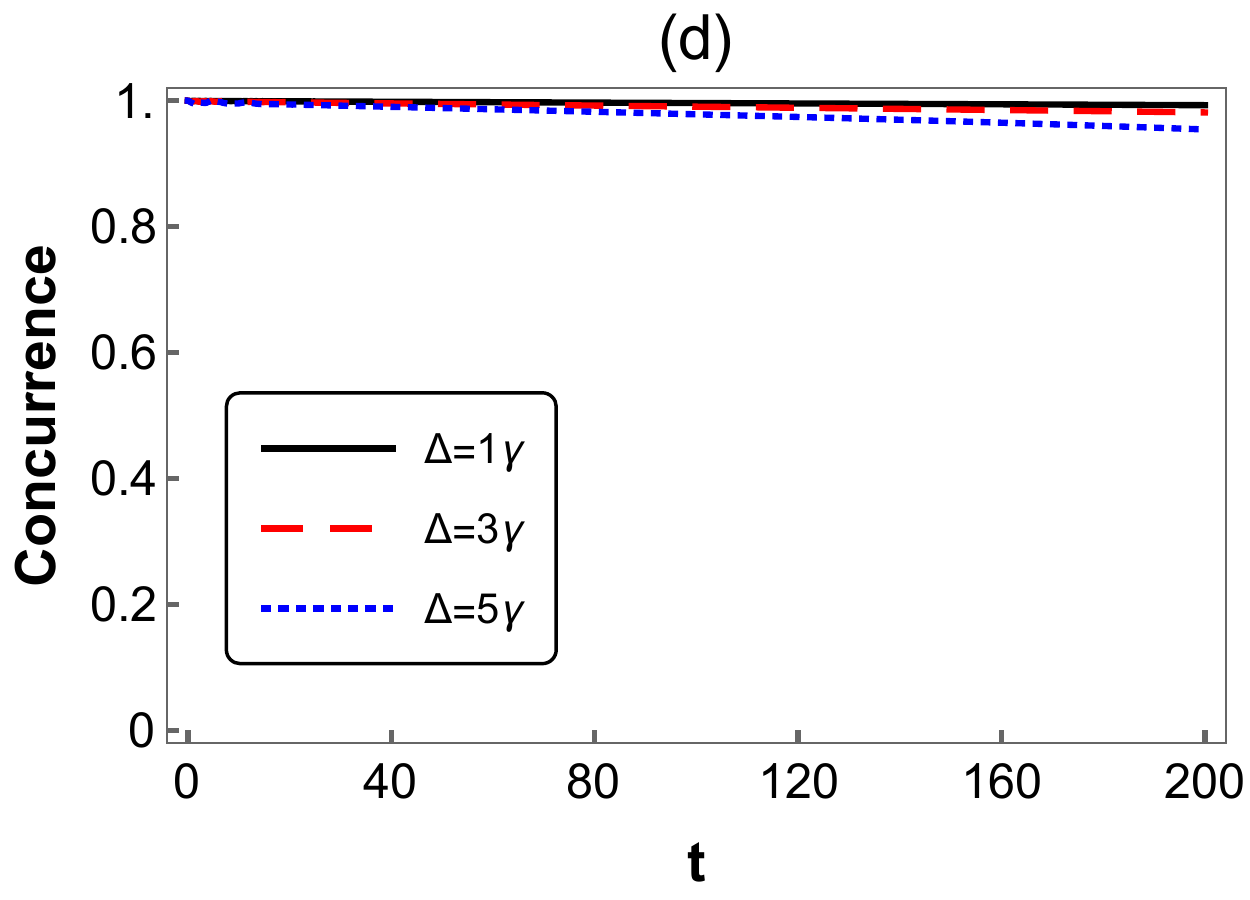}
\caption{Concurrence as a function of time t for different qubits velocity and driving strength under detuning conditions in the strong coupling regime ($\lambda =0.1\gamma$). (a) shows the entanglement evolution in the case of $\Omega=0.5$ and $\beta=0$, (b) shows the entanglement evolution in the case of $\Omega=1.6$ and $\beta=0$, (c) shows the entanglement evolution in the case of $\Omega=0.5$ and $\beta=1 \times 10^{-9}$, (d) shows the entanglement evolution in the case of $\Omega=4$ and $\beta=1 \times 10^{-9}$. Other parameters are $\omega _{0}=1.5 \times 10^{9}$, $\gamma =1$, $\phi=0$, $r_{1}$ =0.5 and $\eta=\frac{\pi}{2}$.}
\label{fig:5}
\end{figure}

Fig. (5) shows the entanglement evolution for various moving-velocity and driving strength under different detuning conditions in the strong coupling regime. Fig. 5(a) shows that the driving strength $\Omega $=0.5 and the velocity ratio $\beta=0$, the entanglement will oscillate and decay rapidly under different detuning conditions. From this picture, we can see that with the detuning increasing, the entanglement will decay faster. In other words, the detuning will have a negative effect on entanglement protection. This is because the dynamics of entanglement are mainly dominated by the interaction between the qubits and the reservoir when the classical driving is very small ($\Omega=0.5$). Fig. 5(b) displays the of entanglement dynamic curves when the driving strength increases to $\Omega =1.6$ and the velocity ratio $\beta =0$. Comparing Fig. 5(b) and Fig. 5(a), we can see that, the amplitude of the entangled oscillation in Fig. 5(b) is smaller than that in Fig. 5(a). And with the driving strength increasing, the decay of entanglement will become obviously slower and the entanglement can survive for a longer time under the detuning case. The reason is that the influence of classical driving on entanglement dynamics will become larger with the driving strength increasing. That is to say, the dynamics of entanglement are determined by both the interaction between the qubits and the reservoir as well as the interaction between the qubits and the driving field when $\Omega=1.6$. Therefore, the negative effects caused by detuning can be suppressed by increasing the driving strength. Fig. 5(c) displays the of entanglement dynamic curves when the driving strength $\Omega $=0.5 and the velocity ratio increases to $\beta =1 \times 10^{-9}$. Compared with Fig. 5(a), we can see that the entanglement decay rate of the moving-qubit system is much smaller than that of the static-qubit system.  Moreover, the entanglement decay rates of the moving-qubit system are approximately equal under the different detuning conditions. This shows that the qubit velocity can reduce the negative effect of detuning on entanglement and can protect entanglement well. Fig. 5(d) shows the of entanglement dynamic curves when the driving strength increases to $\Omega$=4 and the velocity ratio maintains $\beta =1 \times 10^{-9}$. We can know that the entanglement almost always keeps their initial value under the different detuning conditions when the driving strength $\Omega=4$. Namely, the entanglement decay rates are very small and the effect of the large detuning on entanglement dynamics is also very small. Comparing Fig. 5(c) and Fig.5(d), we discover that the entanglement of the moving-biparticle system can also be better protected and the influence of detuning on entanglement is greatly suppressed when the driving strength increases.

Therefore, the different detuning has a great effect on the entanglement of the static-qubit system but little effect on the entanglement of the moving-qubit system. For both the moving-qubit system and the static-qubit systems, the classical driving strength plays an important role in the regulation of entanglement dynamic. The influence of detuning and moving-velocity on entanglement dynamics can be effectively suppressed by the strong classical driving so that the entanglement almost always keeps their initial value.

\section{CONCLUSIONS}
Summary, we investigated the entanglement dynamics of a moving-biparticle system coupled with a zero temperature common environment, in which qubits are driven by an external classical field. The analytical expressions of the density operator and the entanglement can be obtained by using the dressed-state basis when the total excitation number is one. We also studied in detail the effects of the particle-environment coupling $r_{1}$, the spectral width $\lambda$, the velocity ratio $\beta$ and the classical driving $\Omega$ on the entanglement dynamics. The results showed that, in the zero velocity limit, our results match with those of \cite{Nourmandipour} if we take the same resonance case as \cite{Nourmandipour}. All that matters is that we have got some new results when the velocity is not equal to zero. Namely, the proper qubit-environment coupling $r_{1}$ and velocity ratio $\beta$, the smaller spectral width $\lambda$ and the stronger classical driving $\Omega$ can all effectively protect the quantum entanglement of the moving-biparticle system. Especially, the classical driving can effectively eliminate the influence of the qubit velocity and the detuning on the quantum entanglement.

\end{document}